\documentclass[useAMS,usenatbib]{mn2e}

\title[Near-infrared studies of fast nova KT Eri]{Nova KT Eri 2009: Infrared studies of a very fast and small amplitude He/N nova.}

\author[Ashish Raj, D.P.K. Banerjee and N.M. Ashok]{Ashish Raj,$^{}$ D.P.K. Banerjee$^{}$
and N.M. Ashok$^{}$
\thanks{E-mail: ashishr@prl.res.in (AR); orion@prl.res.in (DPKB); ashok@prl.res.in (NMA);} \\
$^{}$Physical Research Laboratory, Navrangpura, Ahmedabad 380009, India\\}

\usepackage[dvips]{graphicx}

\begin{document}

\date{Accepted  Received }

\pagerange{\pageref{firstpage}--\pageref{lastpage}} \pubyear{2013}

\maketitle

\label{firstpage}

\begin{abstract}
We present near-infrared spectroscopic and photometric observations of the nova KT
Eridani taken during the first 100 days following its discovery in 2009 November.
The $JHK$ spectra of the object have been taken from the Mount Abu Infrared
Observatory using the Near-Infrared Imager/Spectrometer. The spectra, typical of
the He/N class novae, show strong He {\sc i} emission lines  together with H {\sc i}
and O {\sc i} emission features. The H {\sc i}, Pa $\beta$ and Br $\gamma$ spectral lines and the He {\sc i} line at 2.0581 ${\rm{\mu}}$m
show broad wings with a relatively narrow central component. The broad
wings extend to $\sim$ 1900 km s$^{-1}$ while the central component has FWHM of $\sim$ 2100
km s$^{-1}$. The $V$ and near-infrared $JHK$ light curves show an additional small amplitude outburst near 40
days after optical maximum. The distance to the nova $d = 6.3$ $\pm$ 0.1 kpc is derived
using the MMRD relation and the estimated value of t$_2$ = 5.7 $\pm$ 0.3 days. The small value of t$_2$
places KT Eri in the class of very fast novae. Using the value of the distance to the nova $d$, we estimate the height of the nova to be
$z = 3.3$ $\pm$ 0.1 kpc below the galactic plane.
We have also calculated  the upper limit for  the ejecta mass for KT Eri to be in the range 2.4 - 7.4 $\times$10$^{-5}$M$_\odot$.
Kinematic evidence is presented from the shape of the line profiles for a
possible bipolar flow. We analyze the temporal evolution of the continuum and also
 discuss the possibility of KT Eri being a recurrent nova.

\end{abstract}


\section{Introduction}

Nova KT Eridani has an interesting  history pre-dating its optical discovery. Its discovery was  reported several days after its actual outburst
 had commenced but in spite of this delay, recourse to archival material helped not only in pinpointing  its exact date of eruption but  also tracked
 its immediate post-outburst evolution as graphically as real-time nova observations usually do. The nova was discovered on 2009 November 25.536 UT
by K. Itagaki at $V$ = 8.1 (Yamaoka \& Itagaki 2009) with a 0.21m telescope patrol system. Itagaki also noted the
presence of a faint object near the position of KT Eri in his archival patrol images. Yamaoka pointed out that the All Sky Automated Survey (ASAS-3)
system (Pojmanski 1997) had detected KT Eri on 3 occasions prior to the discovery announcement of Yamaoka. These pre-discovery magnitudes along
with additional detections are listed in Table 1. In addition Yamaoka suggested that the discovery of KT Eri might be the result of brightening
of a 15th magnitude blue star, listed in many catalogues, with one of the identifications being USNO-B1.0 0798-0048707. Guido et al. (2009)
confirmed the discovery of KT Eri using their CCD observations using a 0.25m reflector and pointed out the presence of a star of $\sim$ 15 mag close
to the position of KT Eri on the DSS red plate of 1990 November 23 UT. The nature of KT Eri as a nova was confirmed by several low resolution
optical spectra obtained during 2009 November 26.5 - 26.6 UT by Fujii, Arai \& Isogai and Imamura (Maehara 2009). These spectra showed broad Balmer
emission (FWHM of H${\rm{\alpha}}$ 3200 - 3400 km s$^{-1}$) together with prominent emission features of He I, N I, N II, Na I D, O I and Mg II
leading Maehara to suggest that KT Eri is a nova of the He/N class.

Rudy et al. (2009) also suggested KT Eri to be a nova of He/N class on the basis
 of their near-infrared spectra obtained on 2009 November 26.4 UT spanning the spectral range of 0.9 to 2.5 ${\rm{\mu}}$m. The He I line at 1.083
${\rm{\mu}}$m was already the strongest line in the spectrum with a P Cygni absorption extending to 3600 km s$^{-1}$. The emission lines of H I, N I
and O I were very strong and broad with FWHM $\sim$ 4000 km s$^{-1}$. The reports of clear detection of KT Eri prior to its discovery has resulted in
 additional studies by different groups to obtain its pre-discovery light curves. Drake et al. (2009) carried out a search of the Catalina Sky Survey
data covering the location of KT Eri during period 2005 January 17 to 2009 November 18. KT Eri was clearly seen in outburst in the images taken on
2009 November 18 and they pointed out that the outburst occurred after 2009 November 10.41 UT. The study of pre-discovery light curve has shown
clear variations of approximately 1.8 magnitudes. The association of an unresolved fainter companion to the 15th magnitude star to be the progenitor
 of KT Eri would mean a very high level of variability. As this possibility is unlikely Drake et al. (2009) conclude that the 15th magnitude star is
most likely associated with KT Eri.

Ragan et al. (2009) derived a distance of $\sim$ 6.5 kpc to KT Eri and pointed out that the 15th
magnitude star located close to the KT Eri position in pre-discovery images is too bright to be its progenitor. McQuillin et al. (2012)
and Hounsell et al. (2010) have obtained an impressive high cadence optical light curve of KT Eri before, during and after the outburst using data from
SuperWASP (Pollacco et al. 2006) and the USAF/NASA Solar Mass Ejection Imager (SMEI) (Jackson et al. 2005) on board the
Coriolis Satellite (Eyles 2003), respectively. The SMEI observations are spread over the period 2009 November 1.13 to 30.62 UT. The light curve, obtained by SMEI
indicates that the initial rise of the nova is steep (rising 4.1 magnitudes over 2.7 days) with evidence of a pre-maximum halt occurring on 2009
November 13.90 $\pm$ 0.04 UT at m$_{SMEI}$ $\sim$ 6. The duration of this halt would seem to be only a few hours, which is appropriate for the
speed of the nova (e.g. Payne-Gaposchkin 1964). The nova reached maximum light on 2009 November 14.67 $\pm$ 0.04 UT with an unfiltered SMEI
apparent magnitude of m$_{SMEI}$ $\sim$ 5.42 $\pm$ 0.02. It subsequently declined rapidly with a $t_2$ value of 6.6 days confirming KT Eri as a
very fast nova. The SMEI observations are consistent with the observed high expansion velocities (FWHM $\sim$ 3400 km s$^{-1}$) and assignment of
He/N spectral class to KT Eri described earlier.

Nesci, Mickaelian \& Rossi (2009) have used the pre-outburst spectrum of the suspected 15th magnitude progenitor
available in the Digitized First Byurakan Survey of the field containing KT Eri to study its nature. The spectra taken on 1971 January 25 shows a
strong UV continuum with several emission lines that include [Ne V] 3426, [Ne III] 3868 and a large blend around H${\rm{\gamma}}$. As these features
are typical of cataclysmic variables, Nesci, Mickaelian \& Rossi (2009) favour the identification of the 15th magnitude star with the actual progenitor of KT Eri
and point out that the forbidden lines of [Ne III] \& [Ne V] were also seen in the transient source CSS 081007: 030559+054715 (HV Cet) suggested to
be a possible He/N nova located at high galactic latitude -43.7 deg (Prieto et al. 2008, Pejcha, Prieto \& Denney 2008). A search for previous outbursts of
KT Eri was done by Jurdana-Sepic et al. (2012) using the Harvard College Observatory archive plates spanning the period 1888 to 1962. As no earlier
outbursts were found, they have suggested that if KT Eri is a recurrent nova, the recurrence time is likely to be on a time scale of centuries.
Jurdana-Sepic et al. (2012) find a periodicity at quiescence of 737 days for the 15th magnitude progenitor that may arise from reflection effects
and/or eclipses in the underlying binary system in KT Eri. KT Eri was detected later as a luminous soft X-ray source (Bode et al. 2010, Ness et al.
2010, Beardmore et al. 2010) and also detected at  radio wavelengths
(O'Brien et al. 2010).

\begin{table}
\begin{center}
\caption{Pre-discovery CCD magnitudes of KT Eri from IAU Circ. 9098.}
\begin{tabular}{cccc}
\hline
Date in UT      & Magnitude   & Observers &   \\
2009 November   &             &   &  \\
\hline
10.236  & $<$ 14.0  & ASAS  &  \\
14.572  &  5.7  & Ootsuki  &  \\
14.632  & 5.4  & Watanabe \& Miyasaita  &  \\
14.813  & 5.6  & Ootsuki \& Ohshima  &  \\
17.226  & 6.9  & Hankey  &  \\
17.758  & 6.7  & Tanaka  &  \\
17.807  & 6.6  & Kawamura  &  \\
18.760  & 7.0  & Yamamoto  &  \\
18.809  & 7.0  & Yamamoto  &  \\
19.241  & 7.34  & ASAS  &  \\
22.179  & 7.98  & ASAS  &  \\
24.269  & 8.12  & ASAS  &  \\
\hline
\end{tabular}
\label{ch5_t1}
\end{center}
\end{table}

\section{Observations}

Near-IR observations were obtained using the the 1.2m telescope of Mt. Abu Infrared Observatory from 2009 November 28 to 2010 March 3. The log of the
spectroscopic and photometric observations is given in Table 2. The spectra were obtained at a resolution of $\sim$ 1000 using a Near-Infrared
Imager/Spectrometer with a 256 $\times$ 256 HgCdTe NICMOS3 array. In each of the $JHK$ bands a set of spectra was taken with the nova off-set to
two different positions along the slit which were subtracted from each other to remove sky and detector dark current contributions. Spectral calibration
was done using the OH sky lines that register with the stellar spectra. The spectra of
the comparison star SAO 131794 (spectral type A3 III;  effective temperature 8600 K) were taken at similar airmass as that of KT Eri to ensure that the
 ratioing process (nova spectrum divided by the
standard star spectrum) removes the telluric features reliably. To avoid artificially generated emission lines in the ratioed spectrum, the H I
absorption lines in the spectra of standard star were removed before ratioing. The ratioed spectra were then multiplied by a
blackbody curve corresponding to the standard star's effective temperature to yield the final spectra.

\begin{table*}
\centering
\caption{Log of the Mt. Abu near-infrared observations of KT Eri. The
date of optical maximum is taken as 2009 November 14.67 UT (Hounsell et al. 2010).}
\begin{tabular}{ccccccccccc}
\hline
Date of      &Days since    &\multicolumn{3}{|c|}{Integration time (s)} &\multicolumn{3}{|c|}{Integration time (s)} &\multicolumn{3}{|c|}{Nova Magnitude} \\
Observation (UT)  &optical maximum             &J-band      &H-band   &K-band  &J-band      &H-band   &K-band   &J-band      &H-band   &K-band  \\

\hline
   &   &\multicolumn{3}{|c|}{Spectroscopic Observations}   &\multicolumn{4}{|l|}{Photometric Observations}  \\
\hline
2009 Nov. 28.854    &14.184&90  &90   &90   &25  &55  &105 &7.39$\pm$0.03 &7.45$\pm$0.03  &7.04$\pm$0.04 \\
2009 Dec. 01.813    &17.143&120  &90   &120  &25  &55  &105 &7.89$\pm$0.03 &7.91$\pm$0.04 &7.37$\pm$0.06 \\
2009 Dec. 03.844    &19.174   &--  &--   &--  &75  &165 &105 &7.88$\pm$0.06 &7.79$\pm$0.04 &7.35$\pm$0.04\\
2009 Dec. 04.834	&20.164    &120  &120   &120  &--  &--  &--  &--&--  &--\\
2009 Dec. 05.844    &21.174&120 &90   &120  &75 &110 &105 &8.21$\pm$0.08 &8.08$\pm$0.04  &7.69$\pm$0.05 \\
2009 Dec. 06.813    &22.143   &90  &90   &120   &--  &--  &-- &-- &--  &-- \\
2009 Dec. 07.865    &23.195 &90  &90   &120   &75 &110 &105 &8.36$\pm$0.01 &8.39$\pm$0.02  &7.95$\pm$0.06\\
2009 Dec. 08.834	&24.164   &120 &90  &120  &--  &--  &--  &-- &--  &--\\
2009 Dec. 09.823    &25.153 &120 &90  &120 &75  &110  &105  &8.39$\pm$0.05 &8.42$\pm$0.07  &8.02$\pm$0.06\\
2009 Dec. 15.776    &31.106  &120 &90  &120 &75 &110  &105  &8.80$\pm$0.03 &8.82$\pm$0.10  &8.35$\pm$0.02\\
2009 Dec. 16.782	&32.112   &120 &90  &120  &--  &--  &-- &-- &--  &-- \\
2009 Dec. 17.786	&33.116   &-- &-- &--  &100  &110  &105 &9.06$\pm$0.06 &8.94$\pm$0.04  &8.55$\pm$0.07\\
2009 Dec. 27.779	&43.109   &-- &--  &--  &100 &220  &105 &9.44$\pm$0.05 &9.34$\pm$0.05  &9.02$\pm$0.13\\
2009 Dec. 29.767	&45.097   &-- &-- &-- &125  &220 &105 &9.53$\pm$0.14 &9.36$\pm$0.03  &8.93$\pm$0.12\\
2009 Dec. 31.757	&47.087   &-- &-- &--  &100  &165 &105 &9.30$\pm$0.06 &9.19$\pm$0.04  &8.85$\pm$0.09\\
2010 Jan. 03.750    &50.080   &-- &--  &--  &250 &550 &105 &9.30$\pm$0.05 &9.91$\pm$0.04  &8.67$\pm$0.07\\
2010 Jan. 25.755	&72.085   &-- &-- &--  &500 &825 &105 &10.58$\pm$0.06 &10.41$\pm$0.05  &9.50$\pm$0.04\\
2010 Jan. 27.744	&74.074   &--&--  &-- &750 &1100 &105 &10.65$\pm$0.02 &10.51$\pm$0.03&9.51$\pm$0.10\\
2010 Feb. 28.735	&106.065   &-- &--  &-- &600 &660 &63 &11.23$\pm$0.08 &11.01$\pm$0.08  &9.99$\pm$0.10\\
2010 Mar. 01.725	&106.790   &-- &-- &-- &600 &660 &63 &11.26$\pm$0.04 &11.05$\pm$0.10  &9.89$\pm$0.10\\
2010 Mar. 03.734	&108.799  &-- &-- &--  &--  &825  &63 &-- &11.05$\pm$0.10  &9.89$\pm$0.10\\

\hline
\end{tabular}
\label{table1}
\end{table*}

Photometry in the $JHK$ bands was done in clear sky conditions using the NICMOS3 array in the imaging mode. Several frames, in 4 dithered positions,
 offset by $\sim$ 30 arcsec were obtained in all the bands. The sky frames, which are subtracted from the nova frames, were generated by median
combining the dithered frames. The star SAO 131500 (spectral type A0, effective temperature 9800K), located close to the nova and
having 2MASS $JHK$ magnitudes 7.35, 7.41 and 7.40, respectively, was used for photometric calibration. Further details of the data
reduction are described in Banerjee \& Ashok (2002). The data were reduced and analyzed using $IRAF$.

\begin{table}
\begin{center}
\caption[List of observed emission lines in $JHK$ spectra of KT Eridani]{A list of the lines identified from the $JHK$ spectra.}
\begin{tabular}{ccc}
\hline\\
Wavelength & Species  &  \\
(${\mu}$m) &   & \\
\hline \\
1.0830   & He\,{\sc i}      &         \\
1.0938   & Pa $\gamma$       &       \\
1.1287          & O\,{\sc i}  &              \\
1.2818   & Pa $\beta$          &   \\
1.5256   & Br 19                &        \\
1.5341   & Br 18                 &       \\
1.5439   & Br 17                  &      \\
1.5557   & Br 16                   &     \\
1.5701   & Br 15                    &    \\
1.5881   & Br 14                     &\\
1.6109   & Br 13                      &  \\
1.6407   & Br 12                       & \\
1.6806   & Br 11                        &\\
1.7002   & He\,{\sc i}                   &\\
1.7362   & Br 10                          &\\
1.9451   & Br 8                            &\\
2.0581 & He\,{\sc i}                        &\\
2.1120,2.1132   & He\,{\sc i}                &\\
2.1656   & Br $\gamma$                        &\\
\hline
\end{tabular}
\label{ch5_t6}
\end{center}
\end{table}

\section{Results}
Before presenting the results proper, we estimate some of the useful parameters of KT Eri like its reddening, distance and height relative to the
 galactic plane from an analysis of its light curve.

\subsection{General characteristics of $V$ and $JHK$ light curves}

The light curves based on the $V$ band data from Table 1 and American Association of Variable Star Observers (AAVSO) and the $JHK$ magnitudes from Mt. Abu
are presented in Fig. 1.
The AAVSO data span the period 2009 November 27 to 2010 March 20. The pre-discovery observations presented in Table 1 from Yamaoka et al. (2009) are also used for
the optical light curve. The epoch of optical maximum for KT Eri is taken as 2009 November 14.67 UT when it reached a value of $V_{max}$ = 5.4. This 
corrosponds to the value of 2009 November 14.67 $\pm$ 0.04 UT obtained by Hounsell et al. (2010) using the SMEI data.

From a least square regression fit to the post maximum light curve we estimate $t_2$ and $t_3$ to be 5.7 $\pm$ 0.3 d and 9 $\pm$ 0.3, respectively,
 making KT Eri
 one of the fast He/N class of novae in recent years. Using the maximum magnitude versus rate of decline (MMRD) relation of della Valle \&
Livio (1995), we determine the absolute magnitude of the nova to be M$_V$ = -8.9 $\pm$ 0.1. The reddening $E(B - V)$ = 0.09, is derived from
Schlegel et al. (1998) towards the nova's direction for which A$_V$ = 0.29 for R = 3.1. Based on the above we obtain a value of the distance
$d$ = 6.3 $\pm$ 0.1 kpc to the nova. Using the value of $d$ we estimated the height $z$ = 3.3 $\pm$ 0.1 kpc below the galactic plane.
 The high galactic latitude is consistent with the low reddening towards the nova. Taking a bolometric correction
of -0.1  appropriate for the SED of novae which resemble  A to F type supergiants near maximum light and the absolute magnitude M$_V$ = -8.9 determined
above gives the the bolometric outburst luminosity of KT Eri to be L$_O$ $\sim$ 3.2 $\times$ 10$^5$ L$_\odot$.

The optical and $JHK$ light curves show an
additional small amplitude outburst near day 40 from the optical maximum. A plateau was also seen in the $V$ band light curve after 80 days from
optical maximum. The duration of the plateau could be tracked for $\sim$ 60 days but its entire duration is difficult to assess as the nova went
into conjunction with the Sun afterwards. The shape of the optical light curve of KT Eri presented in Fig. 1 has all the characteristics
of a P class of nova as per the classification system presented by Strope, Schaefer \& Henden (2010).

The observed value of the outburst amplitude of $\bigtriangleup V$ = 9.6 and $t_{2}$ = 5.7 days for KT Eri is well below the lower limit in the amplitude
versus decline rate plot for classical novae presented by Warner (2008) which shows an expected value of $\bigtriangleup V$ = 12 to 15 for $t_{2}$ = 5.7
 days.
It is also interesting to note that the estimated value of $t_{2}$, $t_{3}$ and $\bigtriangleup V$ are very much similar to those of the
recurrent nova V745 Sco which had corresponding values of 6.2, 9.0 and 9.2, respectively (see Table 17, Schaefer 2010). So, if KT Eri is indeed
a recurrent nova, then the system may be similar to V745 Sco and a possibility exists that the recurrence time for KT Eri may be expected to be around
 50 years similar to the 52yr recurrence time for V745 Sco which had recorded outbursts in 1937 and 1989 (Schaefer 2010).

\begin{figure}
\begin{center}
\includegraphics[width=3.3in,height=3.5in,clip]{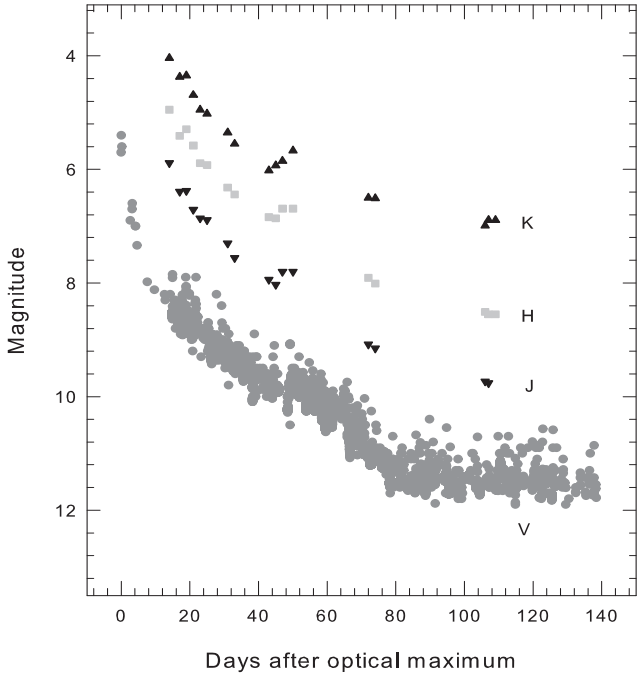}
\caption[The $V$ and $JHK$ band light curves of KT Eri]{The $V$ and $JHK$ band light curves of KT Eri are presented. The $V$ band data is taken from Table 1
\& AAVSO and $JHK$ band data is taken from Mt. Abu observations.}
\label{ch5_2}
\end{center}
\end{figure}

\subsection{Line identification, evolution and general characteristics of the $JHK$ spectra}

The $JHK$ spectra are presented in Figs 2 to 4 respectively; line identification is given in Table 3.
The infrared observations presented here cover the phase after optical maximum with the first infrared spectra taken on 2009 November 28.
In case of He/N class of novae a smaller  number of spectral lines in the near-IR are normally seen
as compared to the Fe II type of novae (Banerjee \& Ashok 2012). Apart from this intrinsic difference between the two nova classes, the possibility
exists of relatively weaker individual lines in He/N novae blending with each other
 due to large expansion velocities and becoming indistinguishable against the underlying continuum. The prominent lines seen in case of KT Eri
are those due to H I, O I and He I. The typical FWHM of the H I lines range from 1900 to 2300 km s$^{-1}$ with full width at zero intensity (FWZI)
 reaching a
value of $\sim$ 4000 km s$^{-1}$. A noticeable feature of these early spectra is the strong presence of lines due to He I. In the spectra taken on 2009
November 28 the He I lines at 1.0830 ${\rm{\mu}}$m, 1.7002 ${\rm{\mu}}$m, 2.0581 ${\rm{\mu}}$m, 2.1120 ${\rm{\mu}}$m and 2.1132 ${\rm{\mu}}$m are
clearly seen. Although it is blended with Pa $\gamma$ at 1.093 ${\rm{\mu}}$m, HeI 1.0830 ${\rm{\mu}}$m is strikingly strong in the spectra. These He I lines are seen to
strengthen and become even stronger as the nova evolves.

No coronal lines are seen  during the course of our observations.
An upper limit can be set for the line flux,  for e.g. for  the [S IX]
1.2520 ${\rm{\mu}}$m line, of 4.6 $\times$ 10$^{-19}$ Wcm$^{-2}$ assuming that a detection could be claimed if the line was present at 5$\sigma$ of the observed continuum noise level
(measured to have a standard deviation of 1$\sigma$) around the expected position of the line. This is based on the later spectra obtained in 2009 December. It is however
difficult to set meaningful upper limits for other coronal lines generally seen in novae spectra because these are either not covered by us (e.g.
lines from [Si VI] 1.964 ${\rm{\mu}}$m, [Si VII] 2.481 ${\rm{\mu}}$m and [Ca VIII] 2.322 ${\rm{\mu}}$m in the K band) or could be
severely blended, if present,  with other lines in the region (e.g. [Al IX] 2.040 ${\rm{\mu}}$m would be blended with the blue wing of the He I 2.0581 ${\rm{\mu}}$m;
Br 10 line at 1.7362 ${\rm{\mu}}$m would blend with [P VIII] at the same position etc). Even the upper limit derived above for [S IX] 1.2520 ${\rm{\mu}}$m, assumes no
contribution from  He I 1.2534 ${\rm{\mu}}$m which occurs at almost the same wavelength.

There is also no indication of dust formation in the nova which is consistent
with the absence of  emission lines from low ionization species like Na and Mg which are potential
 harbingers of dust formation (Das et al. 2008; Banerjee \& Ashok 2012).

\begin{figure*}
 \begin{center}
 \includegraphics [width=5.5in,height=6.0in]{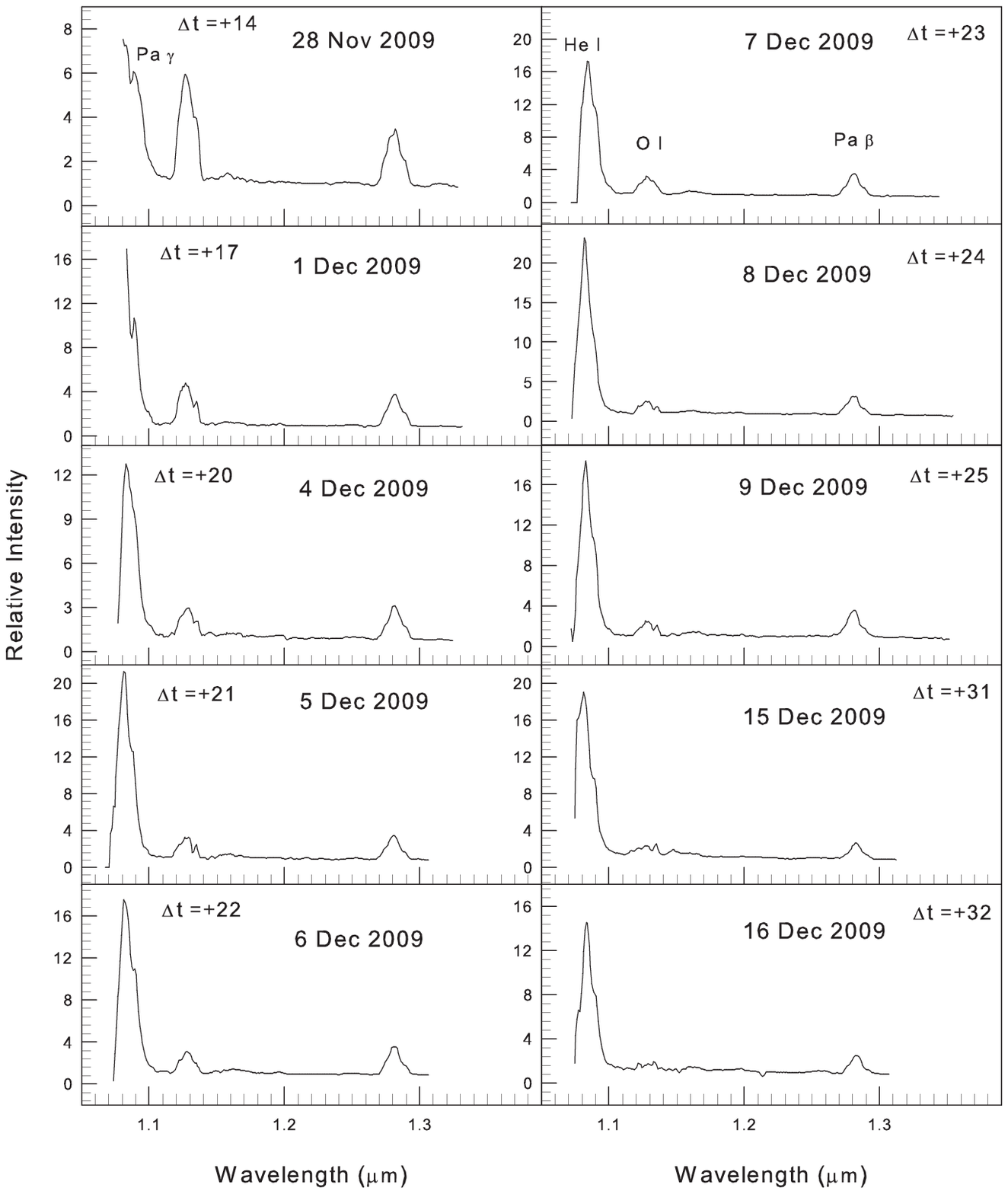}
  \caption[The $J$ band spectra of KT Eri]{The $J$ band spectra of KT Eri are shown at different epochs. The relative intensity is normalized to
unity at 1.25 ${\rm{\mu}}$m. The time $\bigtriangleup$t from optical maximum is given for each spectrum. The continuum flux level
(W cm$^{-2}$ ${\rm{\mu}}$m$^{-1}$) can be calculated for those spectra that have broadband $JHK$ magnitudes for flux calibration.
For such days, the continuum levels in the plot may be obtained by multiplying with 3.8e-16, 2.4e-16, 1.8e-16, 1.5e-16, 1.4e-16 and 1.0e-16 for
the data of 2009 November 28 and 2009 December 1, 5, 7, 9, 15, respectively.}
 \label{ch5_3}
 \end{center}
 \end{figure*}

\begin{figure*}
\begin{center}
 \includegraphics[width=5.5in,height=6.0in]{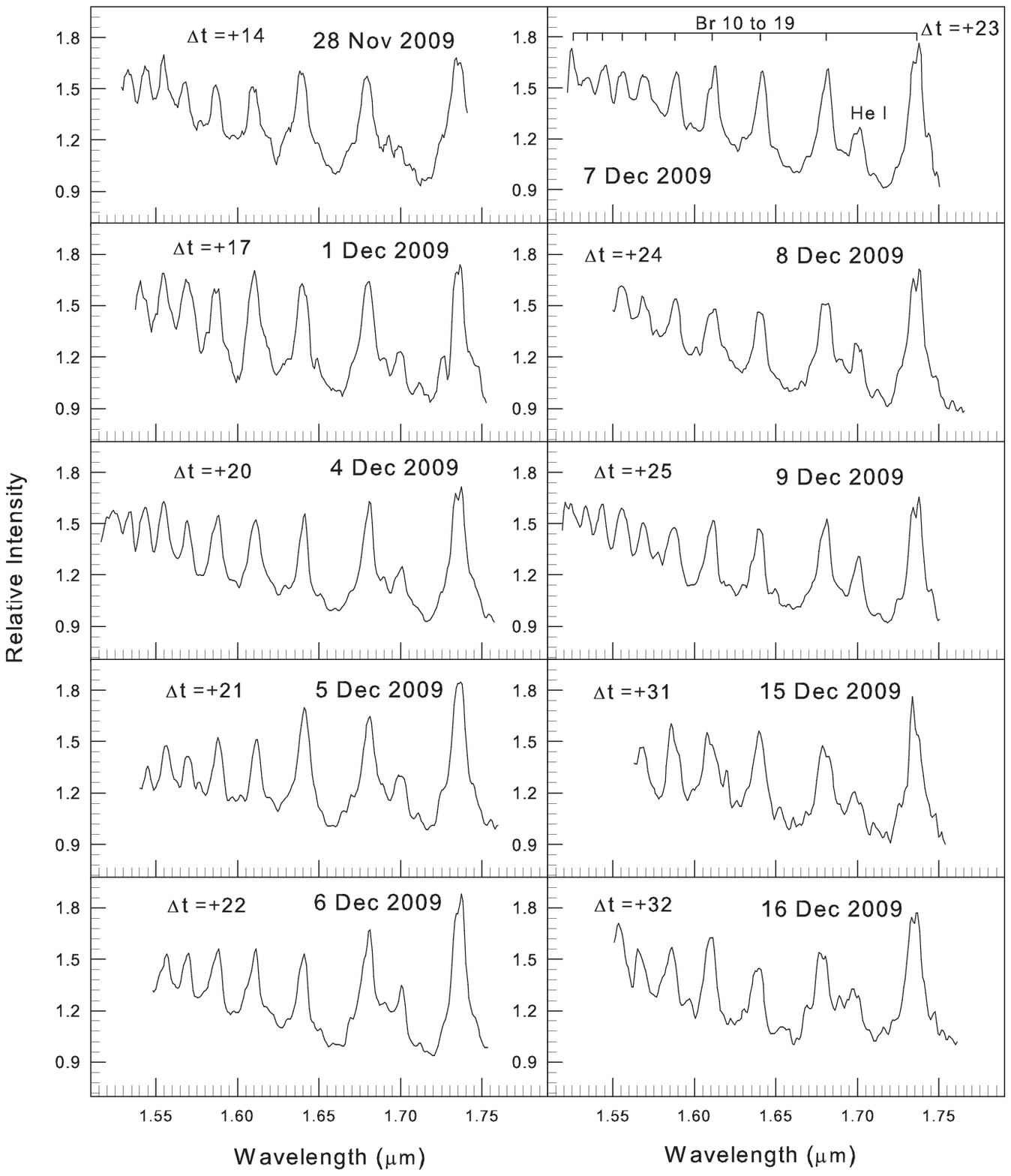}
  \caption[The $H$ band spectra of KT Eri]{The $H$ band spectra of KT Eri are shown at different epochs. The
   relative intensity is normalized to unity at 1.65 ${\rm{\mu}}$m. The time $\bigtriangleup$t from optical maximum is given for each spectrum.
The continuum flux level
(W cm$^{-2}$ ${\rm{\mu}}$m$^{-1}$) can be calculated for those spectra that have broadband $JHK$ magnitudes for flux calibration.
For such days, the continuum levels in the plot may be obtained by multiplying with 1.3e-16, 8.2e-17, 7.0e-17, 5.3e-17, 5.1e-17 and 3.5e-17 for
the data of 2009 November 28 and 2009 December 1, 5, 7, 9, 15, respectively.}
  \label{ch5_4}
  \end{center}
  \end{figure*}

  \begin{figure*}
  \begin{center}
  \includegraphics[width=5.5in,height=6.0in]{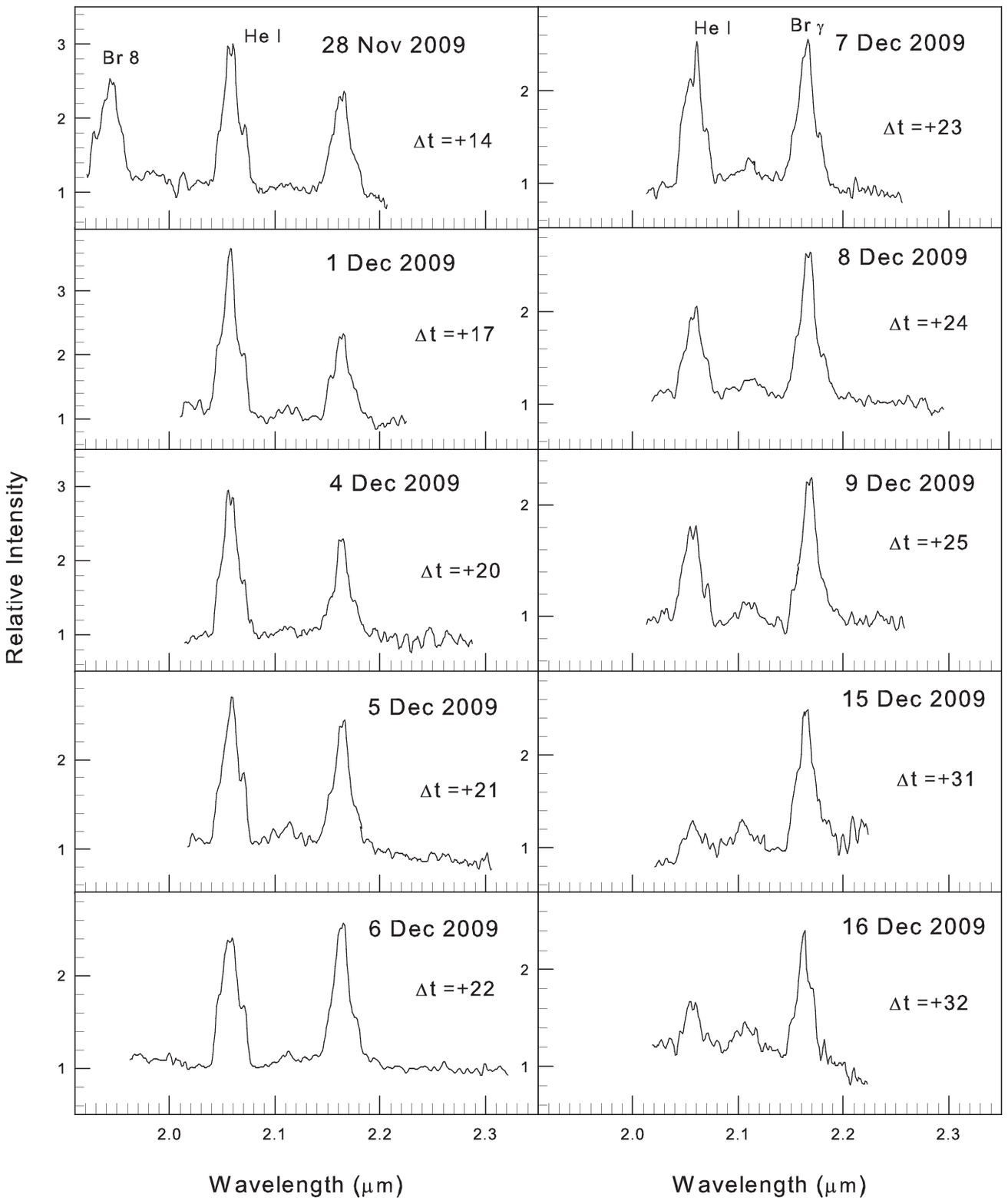}
 \caption[The $K$ band spectra of KT Eri]{The $K$ band spectra of KT Eri are shown at different epochs. The
   relative intensity is normalized to unity at 2.2 ${\rm{\mu}}$m. The time $\bigtriangleup$t from optical maximum is given for each spectrum.
The continuum flux level
(W cm$^{-2}$ ${\rm{\mu}}$m$^{-1}$) can be calculated for those spectra that have broadband $JHK$ magnitudes for flux calibration.
For such days, the continuum levels in the plot may be obtained by multiplying with 6.2e-17, 4.6e-17, 3.4e-17, 2.7e-17, 2.5e-17 and 1.9e-17 for
the data of 2009 November 28 and 2009 December 1, 5, 7, 9, 15, respectively.}
     \label{ch5_5}
     \end{center}
  \end{figure*}

\subsection{Recombination analysis of the H I lines and estimate  of the ejecta mass}

 Recombination case B analysis for the H I lines was done for the observed spectra of all days. Representative results for four days (see Fig. 5)  show the observed 
 strength of the Brackett lines with respect to Br12 which is normalized to unity. Also shown  are the predicted Case B values  from Storey \& Hummer (1995) if the
 lines are optically thin.
The Case B values are for T = 10$^4$ K and for a few representative electron densities of  n$_e$ = 10$^8$, 10$^9$ and 10$^{11}$ cm$^{-3}$.
High electron densities are also considered because the ejecta material may be expected to be dense in the early stages after the outburst. Fig. 5 shows
that the observed line intensities clearly deviate from case B values during the initial observations. Specifically, Br${\rm{\gamma}}$,
which is expected to be relatively stronger,  for any combination of electron density and temperature in the Case B scenario, than the other Br lines considered here
is observed to be considerably weaker in the early observations.  This is certainly
due to large optical depths in the Br gamma line and possibly also in the other Brackett lines (e.g. Lynch et al. 2000).
Such deviations from  case B strengths, early  after outburst, can be expected and have been observed in other
novae too like  V2491 Cyg and V597 Pup (Naik, Banerjee \& Ashok 2009). However,  25 days after maximum  on 2009 December 9 (fourth panel of Fig. 5) there is a clear-cut
indication that case B conditions have begun to prevail implying the lines have become optically thin. For 2009 December 9, it is found that the observed data match well 
with the predicted values for the recombination case B values of T = 10$^4$ K and
an electron density n$_e$ = 10$^9$ cm$^{-3}$. We use additional arguments to constrain the electron density and thereby estimate the mass. 

It is known from Hummer \& Storey (1987) and Storey \& Hummer (1995) that
line center optical depth values can be significant when the electron densities
become large. The value of the opacity factor $\Omega$$_{n,n'}$ for different H{\sc i} lines for transitions between levels $(n,n')$ at different 
temperatures and densities (equation 29 of Hummer \& Storey (1987)) is tabulated in these studies.
From the value of $\Omega$$_{n,n'}$, the optical depth at line-center $\tau$$_{n,n'}$ is given by $\tau$ = $n_e$$n_i$$\Omega$$L$, where $L$ is the path length in cm.
In our case, for the data of 2009 December 9, we take L = 4.84 - 7.26 $\times$ 10$^{14}$ cm corresponding to the kinematic extent of the ejecta using an 
expansion velocity between 2000 - 3000 km s$^{-1}$ (Ribeiro 2011; this work) and a 
time of 28 days since start of outburst. Slight variations in these numbers do not affect our analysis. Since the lines are optically thin on 2009 December 9, this means that
$\tau$ must be less than 1.  For $T_e$ = 10$^4$ K and $n_e$ =
10$^{9}$ cm$^{-3}$, the corresponding value of the
optical depth $\tau$ from Storey \& Hummer (1995) is then found to be 0.016
for Br$\gamma$ and this then decreases monotonically down the series to a
value of $\tau$ = 0.003 for Br 19. At a higher density of $n_e$ = 10$^{10}$ cm$^{-3}$,
the $\tau$ values increase sharply by approximately a factor
of a few hundreds (since $\tau$ $\propto$ $n_en_i$) to $\tau$ $\sim$ 8 for Br$\gamma$ and 0.31 for Br19. Hence, it is seen that 
the Br$\gamma$ line (as also the other Br lines) is/are optically thin at a density of 10$^{9}$ cm$^{-3}$ but becomes thick at 10$^{10}$ cm$^{-3}$.
We therefore adopt $n_e$ = 10$^{9}$ cm$^{-3}$ in our calculations. In applying the relation $\tau$ = $n_e$$n_i$$\Omega$$L$ given by Hummer \& Storey (1987), 
it is assumed that the volume filling factor is unity. This assumption is also carried into the analysis 
below which only aims at setting an upper limit for the ejecta mass.

\begin{figure}
  \begin{center}
\includegraphics[width=3.0in,height=4.0in, clip]{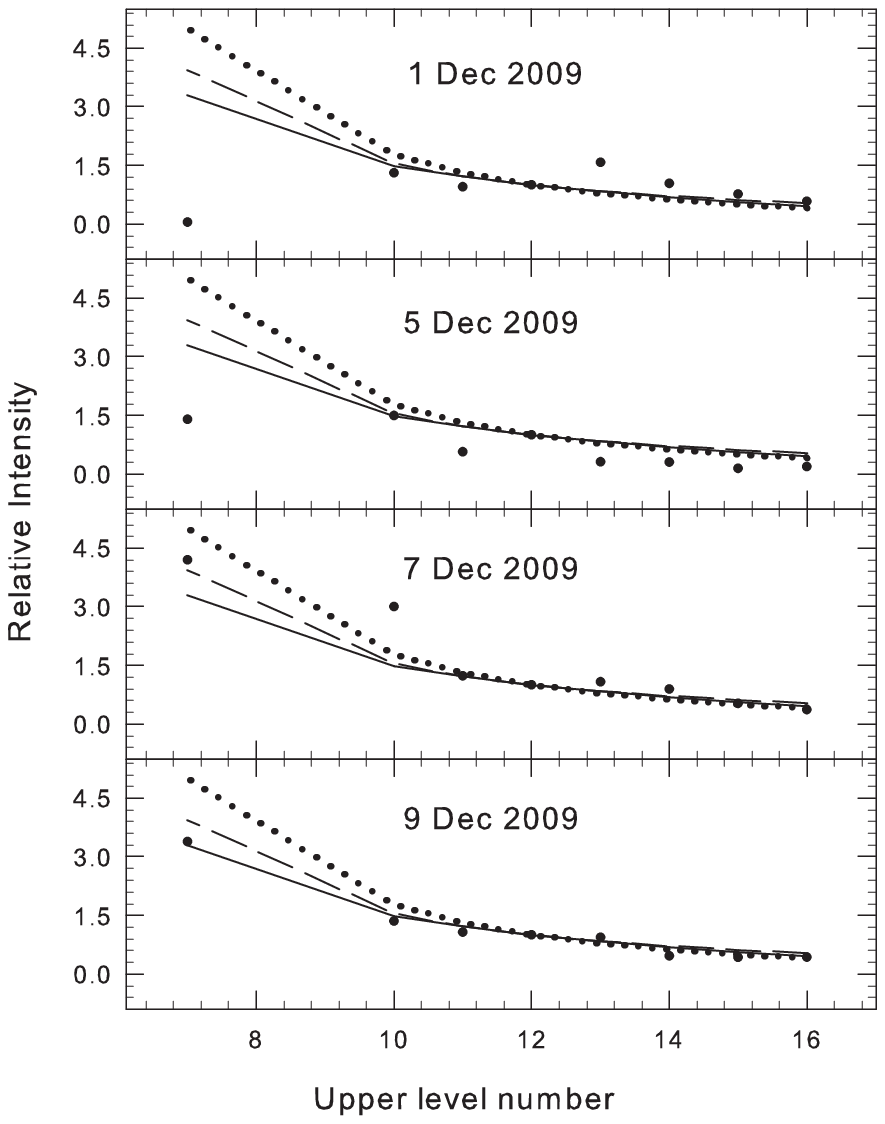}
  \caption[The Case B analysis of H I emission lines]{Recombination analysis for the hydrogen Brackett lines in KT Eri on selected dates.
The abscissa is the upper level number of the Brackett series line transition. The line intensities are relative to that of Br 12. The Case B model
 predictions for the line strengths are also shown for a temperature of T = 10$^{4}$ K and electron densities of n$_e$ = 10$^{11}$ cm$^{-3}$
 (dotted line), 10$^9$ cm$^{-3}$ (solid line) and 10$^8$ cm$^{-3}$ (dashed line).}
  \label{ch5_6}
  \end{center}
  \end{figure}

 Following Banerjee et al. (2010), a constraint on the mass of the emitting gas can be obtained from:

\begin {equation}
M = (4 \pi d^2 (m_H)^2(fV/\epsilon))^{0.5}
\end {equation}
where $d$ is the distance, m$_H$ the proton mass, $f$ the observed flux in a particular line, $\epsilon$ the corresponding case B emissivity;
 V is the volume of the emitting gas, which is [4/3$\pi$ $(vt)^3$ $\phi$], where $\phi$, $v$, and $t$ are the filling factor, velocity and time
after outburst, respectively.
Using the value of temperature and electron density we calculated the mass of the gas in the ejecta in the range 2.4 - 7.4 $\times$10$^{-5}$M$_\odot$.
As the filling factor $\phi$ $<$ 1, this range is an upper limit for the mass of the ejecta.

\subsection{Evidence for a bipolar flow}

A significant finding that has emerged from the analysis of line-profile studies of KT Eri is the likely presence of a bipolar flow in the ejected
material (Ribeiro 2011). Bipolar morphologies are commonly encountered in planetary nebulae and explained on the basis
 of the fast-winds from the hot PN nuclei interacting with a non-uniform
circumstellar environment. In this anisotropic scenario, the pre-existing circumstellar material has a density enhancement in the equatorial plane
 which impedes the  outflowing ejecta
from expanding in the equatorial region while expanding relatively more freely in the polar direction. This leads to a constriction of
the nebula with an hourglass shape. Kinematically, this would imply that the matter in the poles would
flow out with a high velocity relative to the matter in the waist of the hourglass.
It is possible that the secondary, which has colors suggestive of a cool giant (Jurdana-Sepic et al. 2012) has either through mass loss
or a common envelope phase provided the necessary equatorial density contrast needed to create the bipolar morphology (Bond \& Livio 1990).
 It is also becoming increasingly apparent with the aid of high-spatial resolution imaging that hourglass or bipolar morphologies could be commonplace
 in novae shells (Chesneau \& Banerjee 2012). Striking examples of these are seen in the novae V1280 Sco (Chesneau et al. 2012), V445 Pup
(Woudt \& Steeghs 2005) and RS Oph (Bode et al. 2007, Harman et al. 2008).

   The kinematic evidence for a bipolar flow in KT Eri is also seen from our near-IR data. To illustrate this, we present in Fig. 6, representative
profiles of the Pa${\rm{\beta}}$ and Br${\rm{\gamma}}$ lines on 2009 December 7.865 UT (similar analysis was also done for the data of 2009 December 6.813 and 8.834 UT).
As can be seen, the profiles have a strong central component
flanked by two weaker components. In both panels of Fig. 6, we have fitted the profile with three Gaussians viz. a central Gaussian for the core
emission component and two Gaussians for the satellite components in the wings.

\begin{figure}
\begin{center}
\includegraphics[width=3.0in,height=3.0in,clip]{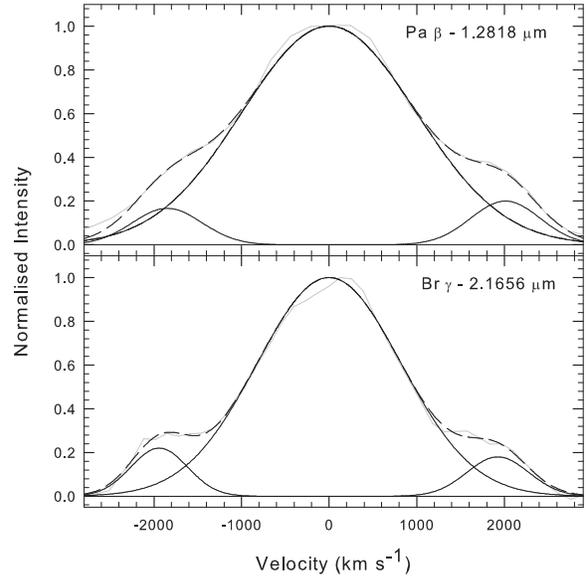}
\caption[Line profiles of Pa${\rm{\beta}}$ and Br${\rm{\gamma}}$ lines]{Line profiles of the Pa${\rm{\beta}}$ and Br${\rm{\gamma}}$ lines on 2009 December
 7 showing broad wings indicative of a bipolar flow. A multi-Gaussian fit of the profiles is shown - a Gaussian for the central component and
two Gaussians for the satellite components in the wings. The Gaussians are shown by continuous black lines, their co-added sum by the dashed line
and the observed data by the gray line. Further details are provided in Section 3.5.}
\label{ch5_7}
\end{center}
\end{figure}

It is seen that a three-component Gaussian fits the data reasonably well. The fits for 2009 December 6, 7 $\&$ 8 indicate the presence of two high velocity components at mean radial
 velocities (averaged over the three days) of -1790 $\pm$ 75 and 2080 $\pm$ 60 km s$^{-1}$ for Pa${\rm{\beta}}$ and at -1800 $\pm$ 115 and 2040 $\pm$ 150 km s$^{-1}$ for the Br${\rm{\gamma}}$ lines, respectively
(these high velocity components appear to be associated with the faster structure reported by Bode et al. (2010)).
The central components have full
width at half-maximum (FWHM) of 2340 $\pm$ 125 and 1930 $\pm$ 85 km s$^{-1}$ for the Pa${\rm{\beta}}$ and Br${\rm{\gamma}}$ lines, respectively. We can interpret the
results of Fig. 6 as follows viz. the core emission can be associated with the slower expanding material from the waist of the bipolar ejecta
while the higher velocity satellite components are associated with the flow from the polar regions. A similar high velocity bipolar flow in the case
of the 2006 outburst of the recurrent nova RS Oph was also deduced by Banerjee, Das \& Ashok (2009) on the basis of broad observed wings in Pa$\beta$ and Br$\gamma$ line profiles.

\subsection{Evolution of the continuum}

We analyze and discuss the evolution of the spectral continuum of KT Eri. At the time of outburst, a nova's continuum is generally well described
by a blackbody distribution from an optically thick pseudo-photosphere corresponding to a stellar spectral type A to F (Gehrz 1988). The spectral
 energy distribution (SED) then gradually evolves into a free-free continuum as the optical depth of the nova ejecta decreases
 (Ennis et al. 1977; Gehrz 1988). The evolution of the continuum of KT Eri is shown in Fig. 7 wherein we have shown representative spectra
sampling the duration of our observations. The spectra in Fig. 7 were flux calibrated using the broad-band $JHK$ photometric observations
presented in Table 2.

\begin{figure}
 \begin{center}
\includegraphics[width=3.0in,height=3.0in, clip]{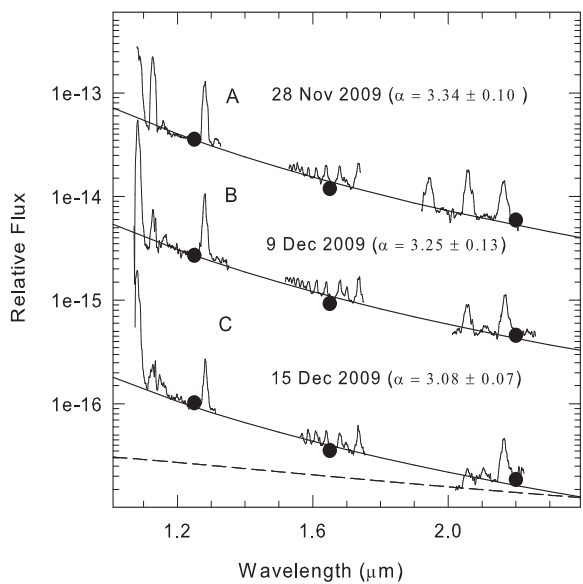}
  \caption[The time evolution of continuum]{The composite $JHK$ spectra of KT Eri for three representative dates 2009 November 28, 2009 December 9
and 2009 December 15 are shown. Model fits to the data with a power law, $F_{\lambda}$ $\alpha$ $\lambda$$^{-\alpha}$, are shown by the continuous
lines with the broad band fluxes represented as filled circles. A decreasing trend in the power law index $\alpha$ is seen indicating the decrease
in the optical depth of the ejecta. The free-free plot (dotted line) for T = 10,000 K is also shown for comparison. }
  \label{ch5_8}
  \end{center}
  \end{figure}

\begin{figure}
\begin{center}
\includegraphics[width=3.0in,height=3.0in, clip]{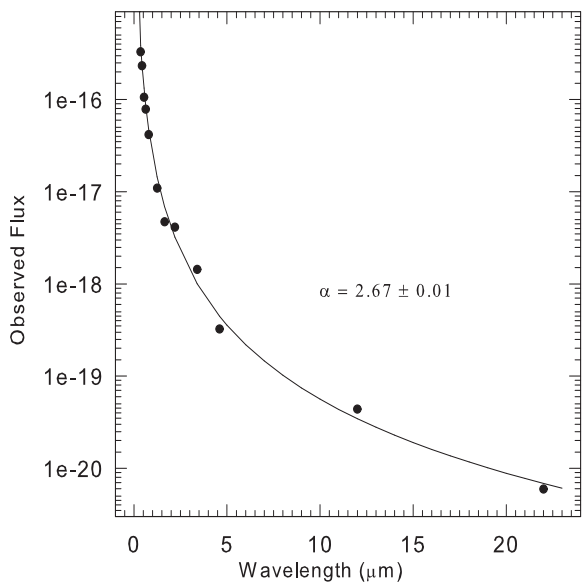}
  \caption[SED during the plateau phase: The power law fit for continuum]{The spectral energy distribution (SED)
 during the plateau phase is shown using $UBVRI$ fluxes in the optical from AAVSO, $JHK$ fluxes in the near-infrared from Mt. Abu observations
 and mid-infrared fluxes from WISE data in units of W cm$^{-2}$ ${\rm{\mu}}$m$^{-1}$ spread over 2010 February 21 to 28. A power law fit with index $\alpha$ = 2.67 is found to fit the data.}
  \label{ch5_9}
  \end{center}
  \end{figure}

At the epoch of optical maximum, a blackbody fit which will have an  index of $\alpha$ = 4.0 at longer wavelengths, is expected to fit the continuum
flux distribution.
As our observations begin from day 14 after optical maximum, we have done power law fits, $F_{\lambda}$ $\alpha$ $\lambda$$^{-\alpha}$,
to see the evolution of continuum spectra shown in Fig. 7.
In the beginning of our observations i.e. on 2009 November 28 (14 d after optical maximum) and 2009 December 9, the continuum spectrum
are well matched with fits having spectral indices of $\alpha$ = 3.34 $\pm$ 0.10 and $\alpha$ = 3.25 $\pm$ 0.13, respectively.
The subsequent spectra taken on 2009 December 15 have become slightly flatter with the slope of $\alpha$ = 3.08 $\pm$ 0.07.
As a comparison, we also show a plot (with dotted line in Fig. 7) of the expected slope for a
case of pure free-free emission evaluated at a representative temperature of T = 10,000 K.

Proceeding further, we have also specifically tried to fit the SED for 2010 February 22 to 28, shown in Fig. 8, because  of the availability of
contemporaneous data at that time including WISE (Wide Field Infrared Survey Explorer) observations (Wright et al. 2010). Such data allows the SED to be fit over a
much larger range in wavelength. In Fig. 8 we have used the $UBVRI$ data from AAVSO, $JHK$ data from Mt. Abu and WISE data in the $W1, W2, W3$ and $W4$
wavebands which are centered at  3.4, 4.6, 12 and 22 ${\rm{\mu}}$m respectively. The corrected flux values corresponding to all these various magnitudes are
 given in Table 4. What we find in Fig. 8 is that  the continuum is remarkably well fit by a power law, with the spectral index $\alpha$ further
reduced  to 2.67. The slow change in the slope of the continuum
spectrum index with time is also evident in the case of other novae, for example, V4643 Sagittarii where it changed from a slope of about 3 to 2 within
about three
months (Ashok et al. 2006). However, in the case of another nova viz. V4633 Sagittarii (Lynch, Russel \& Sitko 2001) the change in the slope of the continuum
was found to be in the opposite direction, i.e. the slope changed from 2 to 2.7 in observations taken 525 and 850 days after outburst. Therefore, the
evolution of the continuum in a nova's development can be fairly complex as also remarked by Lynch, Russel \& Sitko (2001).

It is also worth noting that the SED fit in Fig. 8 represents the behaviour of the nova in the plateau phase. The optical mid-plateau phase has been
clearly seen earlier in a limited number of  novae, notably in the recurrent novae RS Oph, U Sco and CI Aql (Hachisu, Kato $\&$ Luna 2007 and references
therein).  It is usually associated with a super-soft X-ray phase (e.g. in RS Oph as described in Osborne et al. 2011) and is interpreted as being caused
by the accretion disk being irradiated by the central hot white dwarf (Hachisu, Kato $\&$ Luna 2007). The disk is bright as long as the  white dwarf, which is
 rapidly shrinking and becoming hotter at this stage,  is luminous because of a thin hydrogen shell still burning on its surface and it becomes dark
 when the shell-burning ends. The mid-plateau phase is poorly studied  and  a SED covering such a large extent in wavelength, possibly not available
for any other nova, should be useful in interpreting this stage in a nova's development. In the case of a steady-state accretion disk around a white-dwarf,
the continuum radiation from the disk can be described by a $F_{\nu}$ $\alpha$ $\nu$$^{1/3}$ relation (Mayo, Wickramasinghe, Whelan 1980 and references therein) or
equivalently  $F_{\lambda}$ $\alpha$ $\lambda$$^{-\alpha}$ where $\alpha$ = 2.33. The observed spectral index $\alpha$ of the SED is rather close to this
 value. However, the near-IR magnitudes of the suggested progenitor in quiescence indicate the secondary to be an evolved star (discussed later in section 4;
Jurdana-Sepic et al. 2012) so its contribution should dominate over the emission component from the accretion disc.
But it would be interesting to verify, more stringently, what fraction of the observed energy arises due to the irradiated accretion disk
 emitting the reprocessed radiation of the central WD.

\begin{table}
\begin{center}
\caption[List of broad band flux for plateau phase]{List of broad band fluxes corrected for A$_V$ = 0.29 for optical, near-IR \& mid IR data during the plateau phase between 2010 February 22 to 28.}
\begin{tabular}{ccc}
\hline\\
Date of Observation (UT) & Band & Flux (Wcm$^{-2}{\rm{\mu}}$m$^{-1}$)\\
\hline\\
2010 Feb. 26.433 & $U$  & 3.3 $\times$ 10$^{-16}$\\
2010 Feb. 26.428 & $B$  & 2.3 $\times$ 10$^{-16}$\\
2010 Feb. 26.428 &$V$  & 1.1 $\times$ 10$^{-16}$\\
2010 Feb. 26.431 & $R$  & 7.8 $\times$ 10$^{-17}$\\
2010 Feb. 26.432 & $I$  & 4.2 $\times$ 10$^{-17}$\\
2010 Feb. 28.735 & $J$  & 1.1 $\times$ 10$^{-17}$\\
2010 Feb. 28.735 & $H$  & 4.7 $\times$ 10$^{-18}$\\
2010 Feb. 28.735 & $K$  & 4.1 $\times$ 10$^{-18}$\\
2010 Feb. 22.768 & $W1$ & 1.4 $\times$ 10$^{-18}$\\
2010 Feb. 22.768 & $W2$ & 3.2 $\times$ 10$^{-19}$\\
2010 Feb. 22.863 & $W3$ & 4.4 $\times$ 10$^{-20}$\\
2010 Feb. 22.982 & $W4$ & 5.9 $\times$ 10$^{-21}$\\
\hline\\
\end{tabular}
\label{ch5_t5}
\end{center}
\end{table}

\section{Evolution of line profile widths: the recurrent nova possibility}

\begin{figure}
\begin{center}
\includegraphics[width=3.0in,height=3.5in, clip]{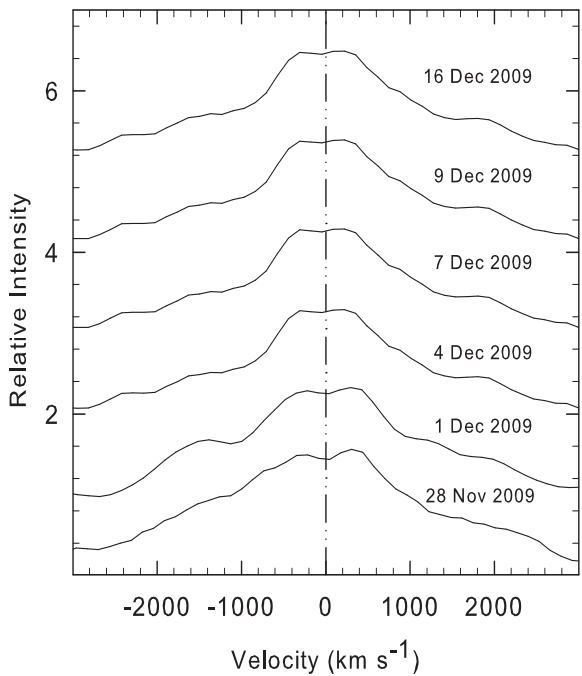}
  \caption[Time evolution of Br${\rm{\gamma}}$ line]{The evolution of Br${\rm{\gamma}}$ line throughout the near-IR observations of KT Eri. The line profiles
  did not change significantly during this period.}
  \label{ch5_10}
  \end{center}
  \end{figure}

The similarity of many of the pre- and post-outburst  UV and optical properties of KT Eri with those of the recurrent novae has raised the question
whether KT Eri could be a recurrent nova (Jurdana-Sepic et al. 2012, Hung, Chen \& Walter 2011, Chen, Hung \& Walter 2013). Jurdana-Sepic et al. (2012)  have analyzed
and listed many of these similarities in details.  There are similarities in X-ray properties too. KT Eri is the second nova, after the recurrent nova
 RS Oph, to exhibit a $\sim$ 35 sec oscillation
in its super soft X-ray emission (Beardmore et al. 2010, Ness et al. 2007, Nelson et al. 2008). Further, if the  bright 15th magnitude star coincident
 with the nova's position is its progenitor, its observed magnitude is suggestive of it being a  bright evolved giant.  It is recognized that of 10 known
recurrent novae, at least eight  are known to harbor evolved secondary stars rather than the main-sequence secondaries typical in classical novae
(Darnley et al. 2012).

The present near-IR observations can, to some extent,  test the suggestion made by Jurdana-Sepic et al. (2012) that the progenitor may be a cool giant star.
These authors  have utilized the 2MASS infrared magnitudes to study the properties of the secondary star in KT Eri as the IR colors are  more sensitive to show
signature of the secondary star. They have calculated the absolute magnitudes of the secondary star to be M$_J$ = 0.48 $\pm$ 0.03, M$_H$ = 0.05 $\pm$ 0.05
and M$_{K_S}$ = 0.00 $\pm$ 0.07 and point out that these magnitudes are  more in line with those of cool giant stars and that the magnitudes  are significantly brighter
than those of typical quiescent CNe (Darnley et al. 2011).

  If KT Eri is indeed a RNe of the RS Oph type, harboring a red-giant companion,  a significant or dramatic decrease in the width of the emission
line profiles may be expected with time after outburst. As the secondary
stars in the RS Oph type RNe lose mass at a high rate, the nova ejecta expelled with high velocity is rapidly  decelerated as it moves through the
wind of
the companion. This deceleration results in a fast temporal decrease  of the expansion velocity of the ejecta as clearly seen  from IR lines in the
 case of the 2006 outburst of RS Oph (e.g Das, Banerjee \& Ashok 2006). We looked for similar behavior in KT Eridani by studying the evolution
of the FWHM of the Br${\rm{\gamma}}$ line.  The line profiles of Br${\rm{\gamma}}$ for 6 days are shown in Fig. 9. These line profiles do not show
 significant changes during the period 2009 November 28 to 2009 December 16 implying a lack of deceleration during this period. There are no observations
in the period between 2009 November 14 to 2009 November 28 to know whether the FWHMs decreased during this period, but it is expected that any deceleration if
 it took place,  would persist for a much longer time (i.e beyond 28 November). This is what was seen in the near-IR in RS Oph (Das, Banerjee \& Ashok 2006) and
also in the optical and near-IR in the symbiotic nova V407 Cyg (Munari et al. 2011) whose secondary consists of a high mass losing Mira variable.

\section{Summary}

We have presented multi-epoch near-infrared spectroscopy and photometry of nova KT Eri which erupted in mid-November 2009. From the optical light curve,
 the distance to the nova, height below the galactic plane and outburst luminosity are estimated.  The light curve classification for KT Eri is seen to
be  type based the plateau observed in its optical and near-IR light curve. The infrared spectra indicate that the nova is of the He/N type. Recombination
 analysis is used to estimate the mass of the gaseous component of the ejecta. Kinematic evidence is presented from the shape of the line profiles for a
possible bipolar flow. We analyze the evolution of the continuum with time and also
 discuss the possibility of KT Eri being a recurrent nova.

\section{Acknowledgments}

The research work at Physical Research Laboratory is funded by the Department of Space, Government of India. This study acknowledges the use of
AAVSO (American Association of Variable Star Observers) and WISE (Wide-field Infrared Survey Explorer) data. We thank the reviewer Prof. A. Evans, for his comments that have helped in improving the results presented here.

\label{lastpage}
\end{document}